\documentclass[letterpaper, 12pt]{article}
\usepackage[margin=1in]{geometry}
\usepackage{setspace}
\usepackage{csquotes}
\usepackage{indentfirst} 
\usepackage{amssymb,amsmath,amsthm,enumitem} 
\usepackage[super]{nth}
\usepackage{threeparttable}
\usepackage{ltablex}
\usepackage{tikz}
\usepackage{float} 
\usepackage{graphicx} 
\usepackage{placeins} 
\usepackage{dcolumn} 
\usepackage{booktabs} 
\heavyrulewidth{0.25ex} 
\usepackage{array} 
\usepackage{placeins}
\usepackage[margin=0.5cm]{caption}
\usepackage{enumitem}
\setlist{nolistsep}
\setlist[enumerate]{topsep=5pt}
\setlist[itemize]{topsep=5pt}
\usepackage{changepage}
\usepackage{rotating}
\usepackage{sectsty} 
\usepackage{setspace} 
\usepackage{titlesec} 
\usepackage{titling}
\newenvironment{my-paragraph}
 {%
  \par
  \setlength{\leftskip}{0cm}
 }
 {\bigskip\par}
\usepackage[sort]{natbib}
\setcitestyle{aysep={}} 
\renewcommand*{\cite}{\citep}
\setcitestyle{notesep={: }}
\usepackage[bottom]{footmisc}
\usepackage{hyperref}
\hypersetup{
    colorlinks=true,
    linkcolor=blue,
    filecolor=blue,      
    urlcolor=blue,
    citecolor=black,
}

\title{The presence of White students and the emergence of Black-White within-school inequalities: \\ two interaction-based mechanisms}

\date{April 12, 2023}

\author{Jo\~{a}o  M. Souto-Maior\thanks{PhD Candidate, Department of Applied Statistics, Social Science, and Humanities---New York University. E-mail: jms1738@nyu.edu; Address: 246 Greene Street (Floor 3), New York, NY 10003, USA.} \\ New York University}

\usepackage{todonotes} 

\begin{document}

\maketitle 

\begin{abstract}
\singlespacing
This article investigates mechanism-based explanations for a well-known empirical pattern in sociology of education, namely, that Black-White unequal access to school resources— defined as advanced coursework—is the highest in racially diverse and majority-White schools. Through an empirically calibrated and validated agent-based model, this study explores the dynamics of two qualitatively informed mechanisms, showing (1) that we have reason to believe that the presence of White students in school can influence the emergence of Black-White advanced enrollment disparities and (2) that such influence can represent another possible explanation for the macro-level pattern of interest. Results contribute to current scholarly accounts of within-school inequalities, shedding light into policy strategies to improve the educational experiences of Black students in racially integrated settings.
\end{abstract}

\pagebreak

\section{Introduction}
The persistence of Black-White educational inequalities in America is an issue of public concern \cite{riegle2018gender}. Because of the known correlation between the level of school resources and the share of White students in the school \cite{reardon2014after, johnson2019children}, it is commonly understood that Black students can benefit from attending schools with a higher presence of White peers, where, presumably, they would have access to more educational resources \cite{frankenberg2007lessons, orfield2005segregation}. However, the presence of White students also correlates with an increase in Black and White students' differential access to educational sources within schools \cite{diamond2006still, tyson2011, mickelson2001subverting, clotfelter2011, lewis2014inequality, lewis2015}. Higher levels of within-school inequality can, therefore, undermine the extent to which Black students benefit from attending schools with a higher share of White peers \cite{diette2021does}.

To shed light into policy strategies that can allow Black students to truly gain the resource-benefits of attending more racially integrated settings, it is, therefore, essential to understand this correlation between the presence of White peers and the distribution of educational resource within schools. To foster this understanding, this paper explores mechanism-based explanations\footnote{Explanations which open the black-box between an explanatory variable and the outcome of interest, by detailing the chain of action-based events linking the two \cite{manzo2022agent, manzo2013educational, hedstrom2005dissecting}.} for an empirical pattern which is representative of this correlation, namely, that Black-White unequal access to a valuable educational resource---advanced coursework---is the highest in racially diverse and majority-White schools \cite{lucas2007race, diette2012whiter, clotfelter2011, tyson2011, kelly2009black}.

\FloatBarrier
\subsection{The empirical pattern of interest}

Curriculum differentiation---the practice of grouping students in different-level courses based on academic background---is a common feature of US public schools \cite{lucas2020race}. This means that students attending the same school often face different curriculum trajectories, with only a few students being able to take the school's most advanced courses---e.g., advanced placement (AP) and honors courses. At least in theory, advanced courses are reserved for students with high academic preparation and motivation \cite{oakes1995matchmaking, kelly2007contours}. In practice, however, placement into advanced courses is a complex process which not only depends on academics, but also a on a series of informal processes---such as the actions of and interactions between school staff, parents and students \cite{tyson2011, lewis2015, oakes1995matchmaking, kelly2007contours, frank2008social}---which can lead to unequal placement decisions even between students with similar academic backgrounds \cite{lucas2020race}.

Empirical investigations show, consistently, that Black and White students' differential access to advanced courses varies across the racial composition of the student body \cite{kelly2009black, tyson2011, lucas2007race, clotfelter2011, diette2012whiter}. Generally, lower gaps are found in majority-Black schools, with higher gaps being observed as the share of White students increase---or when the share of Blacks decrease---\cite{kelly2009black, tyson2011}. Importantly, studies point out that this relationship is not necessarily linear \cite{diette2012whiter, clotfelter2011, lucas2007race}. An increase in the share of Whites is strongly correlated with an increase in White enrollment advantages up until the point where Whites represent about half of the student body, but, after that, the White advantage does not increase as rapidly, and might even start to decrease \cite{diette2012whiter, clotfelter2011}.

This empirical pattern is robust to a series of student-level and school-level controls \cite{kelly2009black, diette2012whiter}. Notably, it is robust to measures of students' prior academic background \cite{kelly2009black}, suggesting that this compositional-dependence arises even between academically comparable students. Further, this relationship is seen after controlling for the share of other race groups in the school \cite{kelly2009black}, and when investigations concentrate on schools where Black and White students represent the majority of the student body \cite{tyson2011}, suggesting that it, to some extent, has to do with the dynamics between Black and White students. 

\begin{figure}[!t]
	\centering
	\singlespacing
\includegraphics{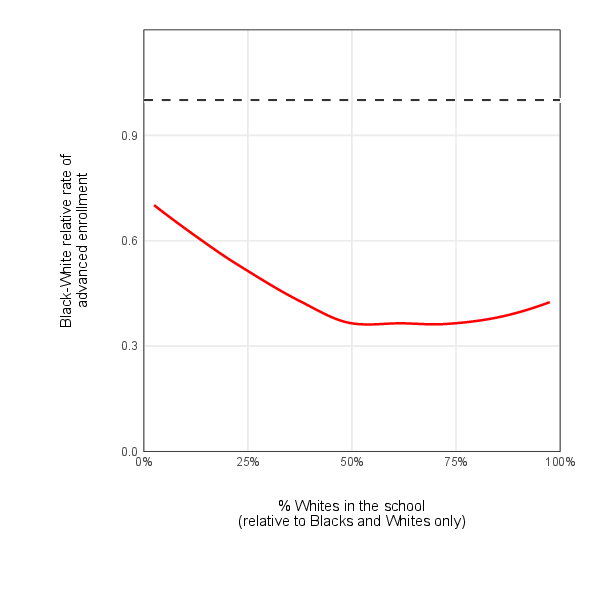}
\caption{\textbf{An empirical relationship between Black-White advanced enrollment inequalities and the presence of White students in school}. Advanced enrollment inequalities are measured as the Black-White relative rate of AP mathematics enrollment in a selected sample of US high schools (see Appendix~\ref{A:sample}). Emphasizing the dynamics between Black and White students, the presence of White students in the school is measured as the number of White students over the number of Black plus White students in the school. Trending line computed by smoothing.}
\label{fig:fig-empirical-comp-pattern}
\end{figure}

A descriptive analysis of Black and White students' enrollment in AP mathematics in US high schools helps illustrating this relationship. Figure~\ref{fig:fig-empirical-comp-pattern} plots the relationship between the Black-White relative ratio of AP mathematics enrollment and the presence of White students in school. A relative rate $< 1$ indicates a White advantage. The closer it is to 0, the higher the White advantage in advanced enrollment. To emphasize the dynamics between Black and White students, the presence of White students in the school is measured as the number of White students over the number of Black plus White students in the school. Appendix~\ref{A:sample} details the sample used in this analysis.

The resulting macro-pattern is captured by a decreasing and concave-up curve, representing two central characteristics of existing empirical findings. First, the White advantage in advanced enrollment tends to increase as the presence of White students in the school increases. Second, this increasing tendency is not linear nor monotonic: the advanced enrollment gap clearly increases up until the share of Blacks and Whites in the school is similar but, after this point, the gap increases at a much slower rate and even starts to decrease when the presence of White students in the school is high.

\FloatBarrier
\subsection{Existing explanations and their limitations}

Although this macro-level relationship is well-established in the literature, studies still struggle to make sense of the mechanisms underlying it. Two common explanations stand out, both of which face important limitations. 

First, authors attribute this pattern to possible racial differences in student peer culture \cite{fryer2010empirical}---which, for reference, I call the \textit{cultural} perspective. This explanation draws on the suggestion that Black students might face social costs for engaging in academic-oriented behaviors---e.g., completing school assignments; striving for good grades; and taking advanced courses---for these actions might be perceived as \enquote{acting White} \cite{fordham1986black}. Importantly, such social costs for \enquote{acting White} seem to be particularly prominent in racially diverse and majority-White schools, but there is no clear indication of such costs in majority-Black schools \cite{fryer2010empirical}. Because such variations in Black peer culture coincides with variations in the Black-White gap in advanced enrollment, the \enquote{acting White} phenomena is often considered as a plausible explanation for the empirical pattern discussed above \cite{lucas2007race, kelly2009black, fryer2010empirical}.

However, this explanation lacks empirical support as there is little evidence of cultural differences in how race groups value academic success \cite{tyson2011, harris2011, lewis2015}. In fact, the correlation between social costs for \enquote{acting White} and majority-White/racially diverse schools is more likely a consequence of Black-White course-taking disparities than an explanation for it \cite{tyson2011, francis2021separate}. The argument is that the strong underrepresentation of Black students in advanced courses often observed in these schools conveys messages which encourage the association of academic success to Whiteness \cite{tyson2011, francis2021separate}.  

This argument, in fact, is the basis for a second common explanation for the pattern of interest---which I call the \textit{structural} perspective. Scholars articulate that when Black students from a given cohort are disproportionally placed in lower-level courses, Black students in the following cohort might fear social isolation in advanced courses, which can reproduce course-taking disparities across racial lines \cite{francis2020isolation}. Then, the historically built underrepresentation of Blacks in advanced courses in majority-White/racially diverse schools can explain why we observe an empirical relationship between school racial composition and Black-White advanced enrollment disparities \cite{tyson2011, francis2021separate}. 



Although this structural view provides a valuable contribution, it only provides an explanation for the \textit{reproduction} (but not \textit{emergence}) of the pattern of interest. In fact, if interpreted as providing a full account of the empirical relationship, it implies that the link between school composition and Black-White advanced enrollment gaps might be, simply, a spurious correlation---where the macro-level pattern is explained by a confounding factor: the historically built underrepresented of Black students in advanced courses in racially diverse/majority-White schools. \textit{In this sense, therefore, it does not consider the extent to which the presence of White students might actually influence the distribution of educational resources between Black and White students.}

With the purpose of complementing this explanation, this paper (1) explores whether we have reason to believe that the presence of White students in school can influence the emergence of Black-White disparities in advanced enrollment and (2) considers whether such influence can represent another possible explanation for macro-level pattern of interest. 

\FloatBarrier
\section{Theoretical background: candidate mechanisms}

Because student placement into advanced courses not only depends on academics, but also on a series of informal, interaction-dependent, processes---i.e., interactions between students, between families and between families and the school \cite{tyson2011, lewis2015, oakes1995matchmaking, kelly2007contours, frank2008social}----, this study considers that the well-known relationship between school racial composition and the tendency of forming same-race ties\footnote{This relationship is defined by two main findings. First, opportunities for cross-race contact in a given school increases with the racial heterogeneity of the student population \cite{blau1977inequality}. Second, \enquote{the observed rate at which cross-race ties are made does not keep up with that opportunity} \cite[p. 703]{moody2001race}, that is, given contact opportunities, homophily---the tendency of forming ties with same-race individuals---also increases with school racial heterogeneity \cite{mouw2006residential, moody2001race, currarini2010identifying}.} \cite{blau1977inequality, mouw2006residential,currarini2010identifying,moody2001race} provides an explanatory path through which the presence of White students in schools can influence Black and White students' access to advanced enrollment. 
In particular, I concentrate on two interaction-based mechanisms which are known to shape students' advanced enrollment chances. These are reviewed below.

\FloatBarrier
\subsection{Information access (IA): the diffusion of information resources}

Navigating the course placement process is a complex task which depends on successful academic strategies and careful planning. To this end, different kinds of information (or knowledge) resources stand out. First, knowledge about how valuable advanced courses can be to one's future social and economic outcomes is essential to shape one's \textit{desire} to take advanced courses \cite{useem1991student, lareau2011, crosnoe2001academic}. Second, information about the schooling processes which define advanced enrollment eligibility\footnote{For example, knowledge about the availability of advanced courses, which course sequence to follow; when to start planning for an advanced curriculum trajectory; which advanced courses best matches the student's academic history \cite{lareau2011, lewis2014inequality, calarco2018}.} and knowledge of how to successfully interact with school staff\footnote{For example, it is known parents, through intensive involvement, can exert pressure on school staff to ensure placement into high-level courses \cite{tyson2011, lewis2014inequality, lewis2015}, and can teach their students to exert such pressure themselves \cite{calarco2018}.} can improve one's \textit{opportunities} for advanced enrollment. 

Parental guidance is often a central avenue through which students gain access to these resources, as parents are important to coordinate and guide students' educational careers and also to mediate how students make sense of the directions given by teachers and school staff  \cite{lareau2011, lewis2014inequality, calarco2018, crosnoe2001academic}. Importantly, some parents, due to their social and educational backgrounds, can have a higher access to such valuable information resources and be better-equipped to assist their children in their educational trajectories \cite{lareau2011, lewis2014inequality}. In addition to gaining access to these information resources through their experiences at home, students (and their families) can also learn such information through social interactions. Students frequently convey their preferences, values and beliefs (often learned at home) with one another \cite{crosnoe2003adolescent, frank2008social} and parents often share information resources in informal and formal environments---e.g., the daily interactions in the community and frequent interactions in parent-teacher organizations \cite{small2009unanticipated, lewis2014inequality}. 

Following the notion that desires (D), beliefs (B) and opportunities (O) shape action (A)---the DBO theory of human behavior \cite{hedstrom2008studying, elster1983sour}---, we can formally describe this mechanism as:
\begin{equation}
\begin{tikzpicture}
    \tikzset{
        force/.style = {rectangle, align=center, font=\footnotesize},
        mylink/.style = {->, dashed}
        }
    \node[force] at (-2,0) (bj) {$B_j$};
    \node[force] at (0,1) (di) {$D_i$};
    \node[force] at (0,0) (bi) {$B_i$};
    \node[force] at (0,-1) (oi) {$O_i$};
    \node[force] at (2,0) (ai) {$A_i$};
    \draw [mylink] (bj)--(bi);
    \draw [mylink] (bi)--(oi);
    \draw [mylink] (bi)--(di);
    \draw [mylink] (di)--(ai);
    \draw [mylink] (oi)--(ai);
\end{tikzpicture}
\end{equation}
This means that that the cognitive beliefs of agent $j$ (their information resources) influence the cognitive beliefs of agent $i$. These beliefs are, then, meaningful to one's desires ($D_i$) and opportunities ($O_i$) for advanced enrollment (action $A_i$).

\FloatBarrier
\subsection{Social belonging expectation (SBE): peer influence on the construction of expectations about social belonging in advanced courses}

Because advanced courses are, generally, elective courses, students are also important actors in the advanced enrollment process. Students' motivation to take advanced courses tends to depend on whether they believe they will \textit{belong} academically and socially in the advanced environment \cite{tyson2011, frank2008social}. For a few students, often the ones accustomed to taking advanced courses since middle school, such beliefs can be \enquote{automatic}, and taking advanced high school courses might feel like a natural next step in their academic trajectory \cite{tyson2011}. For most students, however, beliefs about whether they will belong in the advanced course are highly susceptible to the messages they receive from their peers. In fact, expectations about whether one would have close peers in the advanced course can convey (or fail to convey) a belief that one will belong in the advanced environment \cite{tyson2011, frank2008social, oconnor2011being}.

As above, we can rely on the DBO-theory of human behavior to spell-out this mechanism:
\begin{equation}
\begin{tikzpicture}
    \tikzset{
        force/.style = {rectangle, align=center, font=\footnotesize},
        mylink/.style = {->, dashed}
        }
    \node[force] at (-2,0) (bj) {$A_j$};
    \node[force] at (0,0) (bi) {$B_i$};
    \node[force] at (2,0) (ai) {$A_i$};
    \draw [mylink] (bj)--(bi);
    \draw [mylink] (bi)--(ai);
\end{tikzpicture}
\end{equation}
This means that the action of others ($A_j$)---i.e., plans for enrollment (or actual enrollment) in advanced courses---can influence one's beliefs ($B_i$)---i.e., their expectation about social belonging in the advanced course---and, ultimately, one's course-taking trajectory (action $A_i$).

\FloatBarrier
\section{Analytical approach}

Based on these mechanisms, the task at hand is, therefore, to (1) investigate whether the levels of Black-White course-taking inequalities brought about by the IA and SBE mechanisms vary across the presence of White students in school and (2) explore if these variations can be useful to explain the empirical macro-pattern observed in the literature (and illustrated in Figure~\ref{fig:fig-empirical-comp-pattern}).

To proceed with this analysis, note that dynamics of the IA and SBE mechanisms depend on the social interactions between multiple actors. Investigating the dynamics of complex processes such as these raises important methodological challenges as it implies predicting the macro-level outcomes of micro-level collective behavior---e.g., it raises challenges to common statistical assumptions about the independence of observations and it requires fine-grained measures of socially embedded individual action which are rarely available to the researcher \cite{manzo2022agent, epstein2006generative, hedstrom2005dissecting}. Given that qualitative research already provides a rich knowledge of the role of these mechanisms in shaping advanced enrollment, what is needed is a tool that can help organize our reasoning and facilitate the understanding of how these processes operate under different contextual conditions. Agent-based modeling (ABM) is a computational tool for the analysis of complex systems which serves this exact purpose as it allows scholars to create agents which act according to predefined behavioral rules and to simulate their interaction across different initial conditions \cite{epstein2006generative, manzo2022agent}. 


This study, therefore, constructs an ABM of student enrollment into advanced courses which can allow us to simulate how the inequalities produced by the candidate mechanisms of interest vary when the composition of the school changes. I consider three well-known strategies to strengthen the empirical conclusions of my agent-based simulations \cite{bruch2015agent, manzo2022agent}. First, to support the \textit{theoretical realism} of the ABM, the model is constructed based on the qualitatively informed behavioral rules which capture some of the central processes which define student selection. Second, I \textit{empirically calibrate} agents' social networks according to existing empirical findings. Finally, I \textit{empirically validate} model outcomes by applying the model to a sample of US high schools and by comparing models' predictions to outcomes observed in the data.

\FloatBarrier
\section{The model}  

The model represents the formation of Black-White course-taking inequalities in a given high school and pays particular attention to the role of the IA and SBE mechanisms in the emergence of these inequalities. 

The model represents a high school environment with \textit{n-agents} agents. Each unique agent represents both a given student and their families. To isolate the phenomenon of interest, the high school environment is defined by several idealizations. First, it offers only two types of courses: regular and advanced. There is a maximum enrollment capacity for the advanced course: \textit{pct-advanced-spots}. Second, it has only Black and White students---an idealization that is helpful to emphasize the dynamics between Black and White students. Finally, all students who start in the school proceed to finish high school---i.e., there are no transfers, no dropouts, and no grade retention. 

Each agent $i$ is defined by four variables (Table~\ref{tab:agent-vars}). Variable $\textit{race}_i$, which can be either Black or White.
Variable $\textit{prob-acad-qualification}_i$, which represents the probability that the agent's academic history prior to high school entry---i.e., prior course-taking patterns, course grades and test scores---meets the academic expectations required for advanced enrollment. Variable $\textit{info-access}_i$, which, following the IA mechanism, captures whether the agent has access to the information resources necessary to successfully navigate the advanced enrollment process. It is a dummy variable which equals $1$ if the agent has access to such information resources and $0$ otherwise. Finally, variable $\textit{social-belonging-expectation}_i$, which, following the SBE mechanism, captures agent's expectation of whether they will socially belong in the advanced environment. It is a dummy variable which equals $1$ if such belonging expectation is optimistic and $0$ otherwise.

\begin{table}[!t]
	\footnotesize
	\centering
	\singlespacing
	\caption{Characteristics of agents}
	\label{tab:agent-vars}
\begin{tabular}{p{5cm}p{7cm}wc{3cm}}
\toprule
\addlinespace[0.6em]
Variable & Description & Measure\\
\addlinespace[0.6em]
\midrule
\addlinespace[0.6em]
\multicolumn{3}{l}{\uppercase{Constants}} \\
\addlinespace[0.3em]
\quad $\textit{race}_i$ & Agent's race & Black or White\\
\quad $\textit{prob-acad-qualification}_i$ & Probability that the agent's academic history meets the academic expectations required for advanced enrollment & $[0,1]$\\
\addlinespace[0.6em]
\multicolumn{3}{l}{\uppercase{Dynamic variables}} \\
\addlinespace[0.3em]
\quad $\textit{info-access}_i$ & Whether the agent has access to the information resources needed for advanced enrollment & Dummy\\
\quad $\textit{social-belonging-expectation}_i$ & Whether the agent expects to socially belong in the advanced course & Dummy\\
\addlinespace[0.6em]
\bottomrule
\end{tabular}
\end{table}

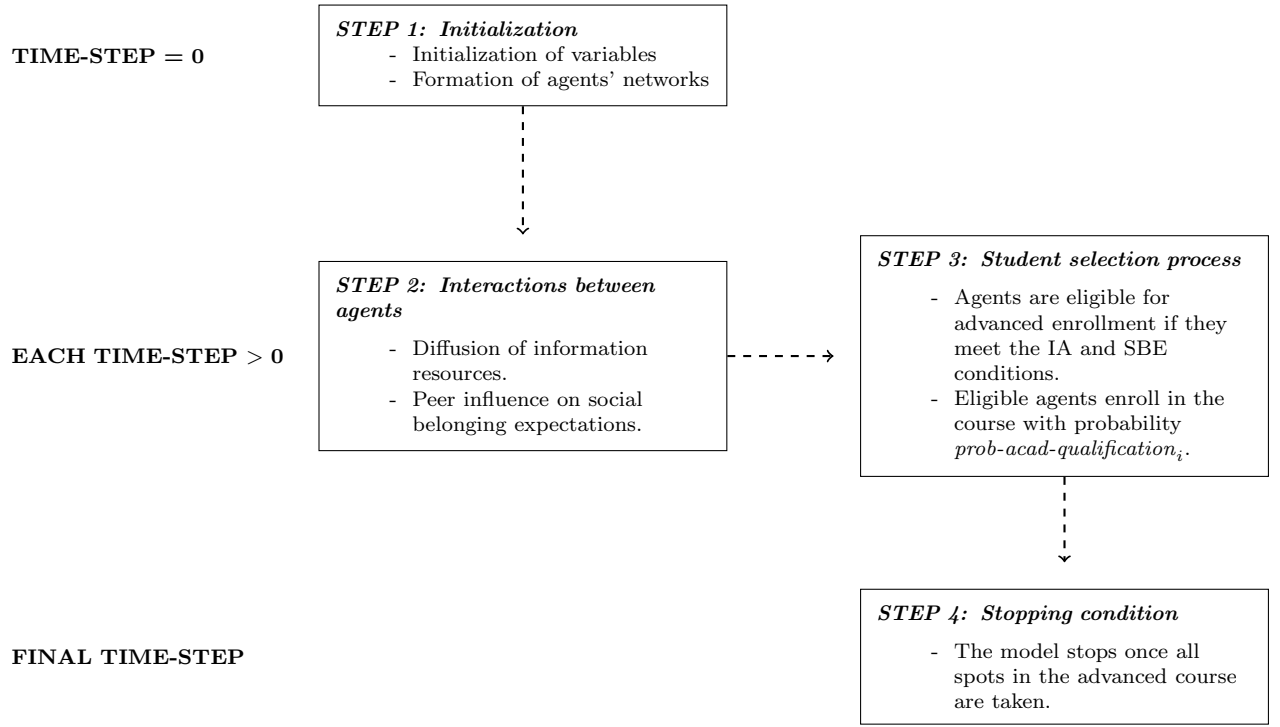
\begin{figure}[!t]
   \scriptsize
\begin{adjustwidth}{0.5cm}{0cm}
\begin{tikzpicture}[x = 6cm, y = 4cm]
	\tikzset{
		comment/.style = {midway},
		level/.style = {rectangle, align=left, inner sep= 6pt, text width=4cm},
		level2/.style = {rectangle, align=left},
		force/.style = {rectangle, draw, align=left, inner sep= 6pt, text width=5cm},
		mymidlink/.style = {-, dashed, thick},
		mylink2/.style = {->, dashed, shorten >=10pt, thick},
		mylink/.style = {->, dashed, shorten >=10pt, thick}}  
	\node[level] at (0,0) (a) {\textbf{TIME-STEP = 0}};
	\node[level] at (0,-1) (b) {\textbf{EACH TIME-STEP $>$ 0}};
	\node[level] at (0,-2) (c) {\textbf{FINAL TIME-STEP}};
	\node[force] at (0.8,0) (1) {\textbf{\textit{STEP 1:}} \textbf{\textit{Initialization}} \\ \begin{itemize}
	                                                                                    \item [-] Initialization of variables
	                                                                                    \item [-] Formation of agents' networks
	                                                   \end{itemize}};
	\node[force] at (0.8,-1) (2) {\textbf{\textit{STEP 2:}} \textbf{\textit{Interactions between agents}}  \\ \begin{itemize}
	                                                                                    \item [-] Diffusion of information resources.
	                                                                                    \item [-] Peer influence on social belonging expectations.
	                                                   \end{itemize}};
	\node[force] at (2,-1) (3) {\textbf{\textit{STEP 3:}} \textbf{\textit{Student selection process}}  \\ \begin{itemize}
	                                                                                    \item [-] Agents are eligible for advanced enrollment if they meet the IA and SBE conditions.
                                                                                     \item [-] Eligible agents enroll in the course with probability $\textit{prob-acad-qualification}_i$.
	                                                   \end{itemize}};
	\node[force] at (2,-2) (4) {\textbf{\textit{STEP 4:}} \textbf{\textit{Stopping condition}}
 \begin{itemize}
 \item [-] The model stops once all spots in the advanced course are taken.
 \end{itemize}};
	\draw [mylink2] (1)--(2);
	\draw [mylink] (2)--(3);
	\draw [mylink] (3)--(4);
\end{tikzpicture}
\end{adjustwidth}
   \caption{\textbf{General structure of the model.}}
   \label{fig:model-process}
\end{figure}

The start of the model represents the beginning of agents' high school years. At time step-step $= 0$, the model initializes its variables and parameters and constructs agents' social networks. After initialization, the model runs in discrete time-steps. Simulation runs capture agents' competition for spots in the advanced course. Each time-step represents an opportunity for social interactions and opportunities for enrollment in the advanced course. Once all spots in the advanced course are taken, the model stops. The final time-step represents the end of agents' high school careers, capturing whether they had the chance to take the advanced course during high school. Figure~\ref{fig:model-process} summarizes the general structure of the model. Computational procedures are discussed in more detail below. 

\FloatBarrier
\subsection{Representation of the student selection process}

Informed by qualitative research, the student selection process is modeled based on three processes. First, is depends on the well-known role of academic background on student selection. Further, it is modelled based on the IA and SBE mechanisms described above.\footnote{A more complete picture might also need to consider that school staff's perceptions of student ability and motivation is subject to racial and class bias \cite{irizarry2015selling, lewis2015}. This notion is not included in the model, first, because the current paper concentrates on inequality-generating mechanisms which are interaction-dependent---and that, because of this dependence, could be influenced by well-known network variations across the racial composition of schools. Second, because the inclusion (or not) of such institutional bias should not influence the dynamics of the IA and SBE mechanism and, thus, for parsimony, it can be omitted.}

These processes are captured in the following computational rule: at each time-step $> 0$, each agent $i$ which is \textit{eligible} (defined later) for the advanced course enrolls in the course with probability $\textit{prob-acad-qualification}_i$. If there are more agents eligible for advanced enrollment than available spots in the advanced course, the model randomly selects only $n$ of these eligible agents for consideration, where $n$ is the number of spots available in the given time-step.

Note that, aligned with qualitative descriptions of the advanced enrollment process, this computational rule imposes that agent's chances of advanced enrollment is directly influenced by their academic background (variable $\textit{prob-acad-qualification}_i$). 

Note, also, that this rule introduces the notion of \textit{eligibility} for advanced enrollment. This notion is introduced as a modeling strategy to capture the IA and SBE mechanisms (it should not be confused with a notion of academic eligibility, something which is already accounted by the fact that agents enroll in the course with probability $\textit{prob-acad-qualification}_i$). Eligibility for advanced enrollment depends on two conditions.

\begin{itemize}
    \item \textbf{The IA condition:} following the IA process, agents are eligible only if they have access to information resources ($\textit{info-access}_i = 1$). 
    \item \textbf{The SBE condition:} following the SBE process, agents are eligible only if they believe they will socially belong in the advanced course ($\textit{social-belonging-expectation}_i = 1$).
\end{itemize}

An important feature of this modeling strategy is that it allows us to separately examine the dynamics of the IA and SBE mechanisms.  In a version of the model (IA version) where eligibility only depends on the IA condition, only the IA mechanism shapes advanced enrollment. Similarly, in a version of the model (SBE version) where eligibility only depends on the SBE condition, only the SBE mechanism shapes advanced enrollment.

\FloatBarrier
\subsection{Dynamics of agent-level variables}

\FloatBarrier
\paragraph{Agent's academic qualification for the course}

Agent's probability of academic qualification for the course is exogenous to the model. To capture the fact that Black and White students often arrive in high school with different levels of academic preparation, I allow $\textit{prob-acad-qualification}_i$ to take one of two values: $\textit{prob-acad-qualification}_w$ for White agents and $\textit{prob-acad-qualification}_b$ for Blacks. The calculation of these probabilities is based on parameter $\alpha_\text{acad-ineq}$, which captures Black's over White's probability of academic qualification for the course. 

To facilitate the modeling of agent's academic background, it is convenient to highlight the role of parameter $\alpha_\text{acad-ineq}$ by de-emphasizing the role played by the absolute values of variables $\textit{prob-acad-qualification}_w$ and $\textit{prob-acad-qualification}_b$. Then, these probabilities are computed in the following way:

\begin{itemize}
    \item If $\alpha_\text{acad-ineq} < 1$ (indicating that Whites arrive in high school with a stronger academic background than Blacks), I set $\textit{prob-acad-qualification}_w = 1$ and $\textit{prob-acad-qualification}_b = \alpha_\text{acad-ineq}$.
    \item If $\alpha_\text{acad-ineq} > 1$ (indicating that Blacks arrive in high school with a stronger academic background than Whites), I set $\textit{prob-acad-qualification}_b = 1$ and $\textit{prob-acad-qualification}_i = 1 / \alpha_\text{acad-ineq}$.
    \item If $\alpha_\text{acad-ineq} = 1$ (indicating that Blacks and Whites arrive in high school with similar academic backgrounds), I set $\textit{prob-acad-qualification}_w = \textit{prob-acad-qualification}_b = 1$.
\end{itemize}

\FloatBarrier
\paragraph{Agent's access to information resources}

According to the qualitative foundations discussed above, parents are a central source of students' access to information resources. Importantly, some parents---often those from upper-class backgrounds---are better-equipped to provide students with such resources. To capture this notion, the model allows some agents to be endowed with unconditional access to information resources (\textit{info-access} $= 1$). The probability of such unconditional endowment depends on the agent's race, being $\textit{prob-unconditional-ia}_w$ for Whites and $\textit{prob-unconditional-ia}_b$ for Blacks, with parameter $\alpha_{ia-ineq}$ determining the level of Black-White inequalities.

Following the IA mechanism, students can also gain access to such resources through social interactions. Then, I define the \textit{diffusion of information resources} procedure: at each time-step, every agent with $\textit{info-access}_i = 1$ randomly selects one of their friendship ties---with $\textit{info-access}_i = 0$---and shares information with probability \textit{prob-diffusion}. Importantly, social interactions are often defined by favoritism towards same-race individuals, even among close ties \cite{aboud2003cross, tilly1998durable}. Then, I allow the probability of diffusion between same-race ties ($\textit{prob-diffusion}_s$) to differ from the probability of diffusion between different-race ties ($\textit{prob-diffusion}_d$), with parameter $\alpha_{\text{favoritism}}$ capturing the level of this difference. For modeling purposes, I set $\textit{prob-diffusion}_s = 1$ and $\textit{prob-diffusion}_d = \alpha_{\text{favoritism}}$.


\FloatBarrier
\paragraph{Agent's construction of social belonging expectations}

Qualitative descriptions point out that for some students---often those which are accustomed to taking advanced courses since middle school---taking advanced courses is simply a natural next step in their academic trajectories \cite{tyson2011, oconnor2011being}. Capturing this possibility, the model allows some students to be endowed with an unconditional belief that they will belong in the advanced course. The probability of such unconditional expectation is $\textit{prob-unconditional-sbe}_w$ for Whites and $\textit{prob-unconditional-sbe}_b$ for Blacks, with parameter $\alpha_{sbe-ineq}$ determining the level of Black-White inequalities.

That said, following the SBE mechanism, such social belonging expectation, for most students, is highly shaped by the messages they receive from their peers. Then, I define the \textit{peer influence on social belonging expectations} procedure: at each time-step, every agent observes the enrollment status of their friendship ties. If at least one of their ties has already chosen to enroll in the course, then the agent believes they will socially belong in the advanced course (sets $\textit{social-belonging-expectation}_i = 1$) with probably \textit{prob-diffusion}. As discussed above, some level of in-group favoritism is often salient in social interactions. Then, I allow the race of one's tie to shape the diffusion of social belonging expectations. If at least one of their same-race ties has chosen to enroll in the course, then the agent sets $\textit{social-belonging-expectation}_i = 1$ with $\textit{prob-diffusion}_s$. In contrast, if no same-race tie has chosen to enroll in the course yet but at least one of their different-race ties has done so, the agent sets  $\textit{social-belonging-expectation}_i = 1$ with $\textit{prob-diffusion}_d$. These probabilities are the same as the ones defined for the \textit{diffusion of information resources} procedure.

\FloatBarrier
\paragraph{Correlation between initial access to information resources and social support}

When both the IA and SBE mechanisms play a role in the model, a correlation between unconditional access to information resources and unconditional access to positive social belonging expectations must be assumed. I define that an unconditional access to information resources is a condition for initial access to positive expectations about belonging.\footnote{Intuitively, if the agent does not have access to the necessary information resources, then the agent might not have the desire nor the opportunity for enrolling in the course, and thus, it does not make sense to say these agents can have an unconditional feeling expectation of belonging in the course.}\footnote{Computationally, among agents initially endowed with $\textit{info-access}_i = 1$, initial endowment with $\textit{social-belonging-expectation}_i = 1$ follows parameter $\alpha_{\text{sbe-ineq}}$. For the group which should have the highest access to $\textit{social-belonging-expectation}_i$, every agent with initial endowment with $\textit{info-access}_i = 1$ is also set to also have $\textit{social-belonging-expectation}_i = 1$. For the other group, only a fraction ($\alpha_{\text{sbe-ineq}}$) of agents with $\textit{info-access}_i = 1$ are also set to also initially have $\textit{social-belonging-expectation}_i = 1$.} 

\FloatBarrier
\subsection{Summary of model variables and parameters}

Beyond the agent-level variables detailed in Table~\ref{tab:agent-vars}, the model description (above) details several parameters and variables which define the model environment and the distribution of agent-level variables. Tables~\ref{tab:parameters} and~\ref{tab:hs-vars} summarize, respectively, these parameters and variables.

\begin{table}[!h]
	\footnotesize
	\centering
	\singlespacing
	\caption{Parameters}
	\label{tab:parameters}
\begin{tabular}{p{3cm}p{12cm}}
		\toprule
		\addlinespace[0.6em]
		Parameter & Description\\
		\addlinespace[0.6em]
		\midrule
		\addlinespace[0.3em]
\addlinespace[0.6em]
\multicolumn{2}{l}{\uppercase{Network structure}} \\
\addlinespace[0.3em]
\quad $\alpha_{\text{homophily}}$ & Fraction of network ties to be with same-race agents\\
\quad $\alpha_{\text{n-ties}}$ & Number of ties to create\\
\addlinespace[0.6em]
\multicolumn{2}{l}{\uppercase{Structural inequalities}} \\
\addlinespace[0.3em]
\quad $\alpha_{\text{acad-ineq}}$ & Black's over White's probability of academic qualification for the advanced course \\
\quad $\alpha_{\text{ia-ineq}}$ & Probability that a Black agent is initially endowed with $\textit{info-access}_i = 1$ over the probability that a White agent is initially endowed with $\textit{info-access}_i = 1$\\
\quad $\alpha_{\text{sbe-ineq}}$ & Probability that a Black agent is initially endowed with $\textit{social-belonging-expectation}_i = 1$ over the probability that a White agent is initially endowed with $\textit{social-belonging-expectation}_i = 1$\\
\addlinespace[0.6em]
\multicolumn{2}{l}{\uppercase{In-group favoritism}} \\
\addlinespace[0.3em]
\quad $\alpha_{\text{favoritism}}$ & Probability of diffusion between different-race agents over the probability of diffusion between same-race agents\\
\addlinespace[0.6em]
\bottomrule
\end{tabular}
\end{table}

\begin{table}[!t]
	\footnotesize
	\centering
	\singlespacing
	\caption{Characteristics of the high school}
	\label{tab:hs-vars}
\begin{tabular}{p{4cm}p{5cm}wc{2cm}wc{5cm}}
		\toprule
		\addlinespace[0.6em]
		Variable & Description & Measure & Default value \\
		\addlinespace[0.6em]
		\midrule
		\addlinespace[0.3em]
\addlinespace[0.6em]
\multicolumn{4}{l}{\uppercase{Constants}} \\
\addlinespace[0.3em]
\quad \textit{n-agents} & Number of agents in the cohort of interest & Integer & From data\\
\quad \textit{pct-available-spots} & The number of available spots in the advanced course over the number of agents & \% &  From data\\
\quad \textit{pct-whites} & Percentage of agents which are White & \% & From data\\
\quad $\textit{prob-unconditional-ia}_w$ & Probability that a White agent is initially endowed with $\textit{info-access}_i = 1$ & $[0,1]$ & From data\\
\quad $\textit{prob-unconditional-sbe}_w$ & Probability that a White agent is initially endowed with $\textit{social-belonging-expectation}_i = 1$ & $[0,1]$ & From data\\ 
\quad $\textit{prob-diffusion}_s$ & Probability of diffusion between same-race agents & $[0,1]$ & 1\\
\addlinespace[0.3em]
\multicolumn{4}{l}{\uppercase{Calculated from constants and parameters}} \\
\addlinespace[0.3em]
\quad \textit{pct-blacks} & Percentage of agents which are Black & \% & $100 - \textit{pct-whites}$\\
\quad $\textit{prob-unconditional-ia}_b$  & Probability that a Black agent is initially endowed with $\textit{info-access}_i = 1$ & $[0,1]$ & $\alpha_{\text{ia-ineq}} \cdot \textit{prob-unconditional-ia}_w$ \\
\quad $\textit{prob-unconditional-sbe}_b$  & Probability that a Black agent is initially endowed with $\textit{social-belonging-expectation}_i = 1$ & $[0,1]$ & $\alpha_{\text{sbe-ineq}} \cdot \textit{prob-unconditional-sbe}_w$\\ 
\quad $\textit{prob-diffusion}_d$ & Probability of diffusion between different-race agents & $[0,1]$ & $\alpha_{\text{favoritism}} \cdot \textit{prob-diffusion}_s$ \\ 
\addlinespace[0.3em]
\bottomrule
\end{tabular}
\end{table}

\FloatBarrier
\subsection{Empirical calibration of agents' social networks}
\FloatBarrier

Because this study's hypothesis depends on the well-known relationship between relative group size and the probability of forming same-race ties \cite{mouw2006residential, moody2001race, currarini2010identifying}, I calibrate the model's formation of agents' networks based on a study that is representative of these findings: \citet{currarini2010identifying}. \ref{A:net} provides a more detailed discussion of this choice. 


Let $r$ indicate the race of agent $i$ and let $r = b$ and $r = w$ indicate, respectively, a Black and a White agent. Also, for a given school, let $G_r$ denote the relative group size of agents of race $r$ in the school; $N_r$ to be the average number of undirected ties for agents of race $r$; and $Q_r$ to be the share of ties formed by agents of race $r$ which are with same-race agents. Then, the relationship between the simulated fraction of same-race ties, $Q_r$, and the relative group size of agents in the model environment, $G_r$, should mimic the empirical findings presented by \citet{currarini2010identifying}---between $G^{true}_{r}$ and $N^{true}_{r}$. An important constraint is that the simulated relationship between $N_r$ and $G_r$ must also follow empirical results ($G^{true}_{r}$ and $N^{true}_{r}$) so that the simulated pattern is not driven by unrealistic changes in the total number of friendship ties.

The target empirical findings \cite{currarini2010identifying} are summarized by the following equations:\footnote{Note that the calibration here simplifies minor Black-White differences in the relationship between $N_r$ and $G_r$ estimated by \citet{currarini2010identifying}. Here, because of the focus on only two race groups, and because agents must have a meaningful share of different-race ties (Equation~\ref{eq:IH}), the characteristics of the networks for the two race groups tends to be strongly dependent on one another. Then, it strikes me as reasonable to calibrate networks based on the aggregated patterns found in their study.}
\begin{align}
N_r^{true} &= 5.54 + 2.27 G^{true}_{r} \label{eq:N} \\
IH_{r}^{true} &= 0.032 + 2.15 G^{true}_{r} - 2.35 (G^{true}_{r})^2 \label{eq:IH}
\end{align}
Equation~\ref{eq:N} estimates a linear, increasing relationship between own-type students in the school and the average number of friendship ties. Equation~\ref{eq:IH} defines Coleman's inbreeding homophily index for race $r$, a measure which compares how homophilius students in the school are compared to how homophilius they could be given contact opportunity, computed by $IH_r = (Q_r - G_r)/(1 - G_r)$. 

To derive the empirical relationship between $Q_r$ and $G_r$, I solve Equation~\ref{eq:IH} for $Q_r$:
\begin{align}
Q^{true}_{r} &= IH_{r}  (1 - G^{true}_{r}) + G^{true}_{r} \nonumber\\ 
    &= (0.032 + 2.15 G_r^{true} - 2.35 (G_r^{true})^2 )(1- G_r^{true}) + G^{true}_{r}\label{eq:Q}
\end{align}
To mimic this empirical pattern, the network formation model follows a simple behavioral rule: each agent $i$ creates a total of $\alpha_{\text{n-ties}, i}$ undirected friendship ties under the condition that a fraction of these ties, $\alpha_{\text{homophily}, i}$, should be with agents of own-type. Parameter $\alpha_{\text{n-ties}, i}$ depends on $G_r$ and varies across each student $i$. It can be written as $\alpha_{\text{n-ties}, i}(G_r)$. Parameter $\alpha_{\text{homophily}, i}$ also depends on $G_r$, but is constant across individuals. Then, it can be written as $\alpha_{\text{homophily}}(G_r)$. The network formation model, therefore, has two parameters of interest: $\alpha_{\text{homophily}}(G_r)$ and $\alpha_{\text{n-ties}, i}(G_r)$ (see Table~\ref{tab:parameters}).

\begin{figure}[!t]
	\centering
	\includegraphics{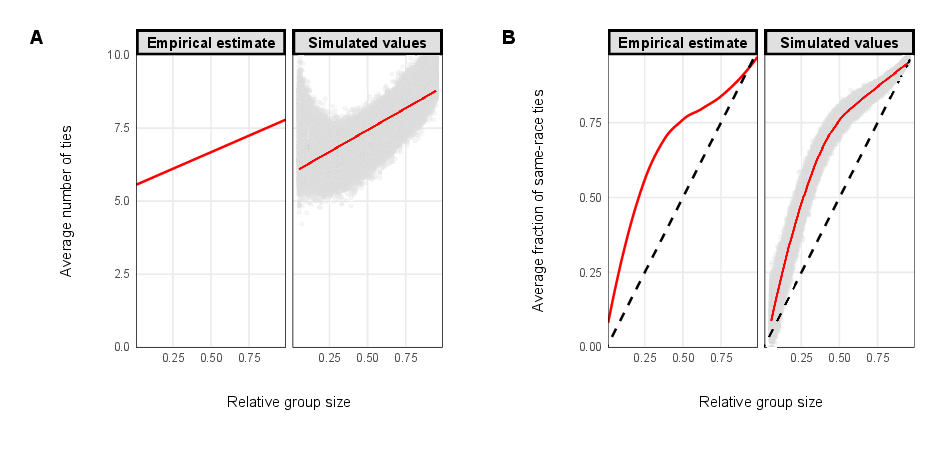}
	\caption{\textbf{Empirical and simulated relationships between relative group size and: agents' average number of friendship ties (Plot A); agents' average fraction of same-race ties (Plot B).} The \enquote{Empirical estimate} panel plots results from \citet{currarini2010identifying} and the \enquote{Simulated values} panel plots results from model simulations. Simulated values result from 1,000 model runs for each integer share of Whites between 1\%-99\%. Trending lines computed by linear regression---Plot A---and smoothing---Plot B.}
	\label{fig:fig-net}
\end{figure}

I define $\alpha_{\text{n-ties}, i}(G_r)$ to be uniformly distributed across agents as follows:
\begin{align}
\alpha_{\text{n-ties}, i}(G_r) \sim U[0, \theta_{0} + \theta_{1} G_r] \label{eq:n-ties}
\end{align}
Note that, following the empirical pattern defined in Equation~\ref{eq:N}, the maximum number of ties an agent can form ($\theta_{0} + \theta_{1}G_r$) varies across relative group size, $G_r$. Model explorations over several combinations $\theta_{0}$ and $\theta_{1}$ show that $\theta_{0} = 5$ and $\theta_{1} = 6$ allow the model to produce a pattern that closely approximates the empirical target (see Figure~\ref{fig:fig-net}A).

Further, I calibrate parameter $\alpha_{\text{homophily}}(G_r)$ according to Equation~\ref{eq:Q}:
\begin{align}
\alpha_{\text{homophily}}(G_r) = (0.032 + 2.15 G_r - 2.35 G^{2}_r )  (1 - G_r) + G_r\label{eq:homophily}
\end{align}
Figure~\ref{fig:fig-net}B shows that the relationship between relative group size and the tendency of forming same-race ties produced by the model closely resembles the empirical patterns of interest. 

\FloatBarrier
\subsection{Empirical initialization}

To provide model variables and parameters with some empirical realism, I apply the model to the context of advanced placement (AP) mathematics---a selective set of high school courses which provide highly-prepared students with college-level mathematics content \cite{hacsi2004document, judson2015growth}. I initialize the model according to data from a sample of 530 US public high schools. \ref{A:sample} details the construction of the sample and discusses how I map empirical variables into model variables. Parameter $\alpha_{\text{favoritism}}$, which is set to 0.9, is an exception to this empirical validation as it chosen based on a sensitivity analysis (\ref{A:sens}).

\FloatBarrier
\subsection{Empirical validation of the model}

To further strengthen the reliability of the ABM constructed here, I also validate simulated outcomes with the empirical application of interest. For each $j$ high school, I calculate an empirical ($D^{true}_{j}$) and simulated ($D^{sim}_{j}$) measure of the White-Black advanced enrollment rate gap, $D_j$.\footnote{Letting $N_b$ and $N_w$ define, respectively, the number of Black and White agents in school and $N^{enroll}_{b}$ and $N^{enroll}_{w}$ define, respectively, the number of Black and White agents enrolled in the schools' advanced course, the White-Black enrollment rate gap for school $j$ can be written as $D_j = 100 (N^{enroll}_{w}/N_w - N^{enroll}_b/N_b)$.} The simulated measure for each high school  ($D^{sim}_{j}$) is computed as the averaged outcome across 100 simulation runs for each high school. Then, I calculate a measure of model performance: \textit{ME}---model's mean error computed as the difference between $D^{sim}_{j}$ and $D^{true}_{j}$ averaged across all $j$ high schools. 

\begin{figure}[H]
	\centering
	\singlespacing
\includegraphics{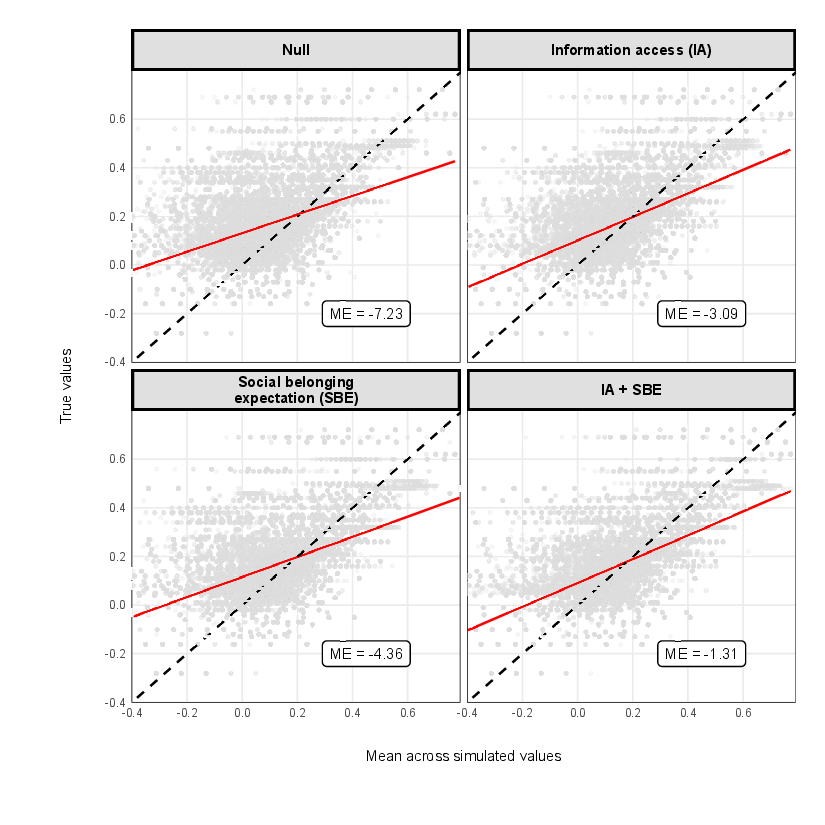}
\caption{\textbf{Empirical validation of the model.} The Figure plots the true (y-axis) and simulated (x-axis) values of White-Black advanced enrollment rate gap in each of the selected 530 high schools (\ref{A:sample}). Results are presented for four different versions of the model: Null; IA; SBE; and IA + SBE, where the version name indicates the mechanisms shaping student placement. In the Null version, placement is only based on academic background. The Figure also reports the model's mean error rate \textit{ME}---the average difference between the true and simulated values across all high schools---and a red regression line which summarizes the tendency of the model's predictions.}
	\label{fig:fig-model-validation}
\end{figure}

I perform such validation procedure across four different versions of the model: Null; IA; SBE; IA + SBE; where the version name indicates the generative processes determining the student selection process. The Null version does not include any of the mechanisms of interest and, thus, models enrollment solely based on differences in academic qualification. Figure~\ref{fig:fig-model-validation} details the results. 


Figure~\ref{fig:fig-model-validation} plots the true ($D^{true}_{j}$, y-axis) and simulated ($D^{sim}_{j}$, x-axis) measures of Black-White AP mathematics enrollment disparities for each of the selected 530 high schools (see \ref{A:sample}). A regression line, in red---which is computed from a regression of the true on all simulated values---summarizes the tendency of the model's predictions. 

I evaluate these results according to three principles. First, because the literature suggests that the ID and SBE processes both shape Black-White disparities, model performance should increase as additional generative processes are included in the ABM, and, thus, the absolute value of \textit{ME} should decrease. Second, the magnitude of the absolute value of \textit{ME} across models---particularly models which include the generative processes of interest---should be reasonably low, indicating that the model reasonably represents an empirical reality. Finally, the model should, on average, tend to underestimate the level of White-Black enrollment gaps. This underestimation is desirable because the model here only includes a subset of the many processes which can shape Black-White advanced enrollment inequalities---e.g., it assumes away the possibility of institutional bias against Black students \cite{irizarry2015selling, lewis2015}. Providing an empirical support to the model constructed here, Figure~\ref{fig:fig-model-validation} shows that the model satisfies these three principles.

\FloatBarrier
\section{Results}
\label{sec:results}

Now, assuming that the model captures the information access (IA) and expectation about social belonging (SBE) mechanisms reasonably well---an assumption supported by qualitatively informed behavioral rules; the empirical calibration of agents' networks; and empirical validation of model outcomes---we can apply it to better understand how the presence of White students in school might influence the emergence of Black-White advanced enrollment inequalities.


To this end, I run a series of model simulations where I allow the share of White agents in the high school to vary while keeping other initial characteristics constant. In these investigations, the outcome of interest is the simulated Black-White relative rate of enrollment in each high school, $RR_{j}$.\footnote{Letting, $N_b$ and $N_w$ to be, respectively, the number of Black and White agents in school and $N^{enroll}_{b}$ and $N^{enroll}_{w}$ to be, respectively, the number of Black and White agents enrolled in the schools' advanced course, the Black-White relative rate of enrollment for school $j$ is $RR_j = (N^{enroll}_{b}/N_b) / (N^{enroll}_w/N_w)$.} 
To better illustrate the dynamics of the model, I discuss four different model versions: Null; IA; SBE; IA + SBE; where the version name indicates the generative processes determining the student selection process. The Null version models student selection only based on differences in academic qualification. 

As outlined in the model description, above, network parameters are calibrated based on existing research; the in-group favoritism parameter is chosen based on a sensitivity analysis (\ref{A:sens}); and other parameters/variables are initialized based on the empirical context of AP mathematics course-taking (as detailed in \ref{A:sample}). In the analysis that follow, to allow a manageable, yet informative, set of simulation results, I initialize the model based on the mean values of the empirical sample of interest (see Table~\ref{tab:data-sources}).\footnote{Variable $\textit{n-agents}$ is an exception. I assume $\textit{n-agents} = 500$ (and not its empirical mean value of 252) because, when exploring models with small fractions of Black and White agents, if the total number of agents is small, model initialization might not work appropriately.} Table~\ref{tab:initial-values} summarizes the initial values for the different model simulations.

Note that because the link between school racial composition and Black-White advanced enrollment gaps reported in the literature is robust to controls for measures of students' academic history, I allow simulations to concentrate on Black and White students with similar academic backgrounds by setting $\alpha_{\text{acad-ineq}} = 1$. 

\begin{table}[!t]
	\footnotesize
	\centering
	\caption{\textbf{Combinations of initial values for variables and parameters for model initialization.} Values marked with $^*$ are defined manually. All other values are initialized with the mean values of the empirical sample of interest (Table~\ref{tab:data-sources}) }
	\label{tab:initial-values}
	\begin{tabular}{l | c c c c}
\hline
\addlinespace[0.3em]
 &  $C_1$  & $C_2$ &  $C_3$ &  $C_4$ \\
 \addlinespace[0.3em]
 \hline
 \addlinespace[0.6em]
\multicolumn{5}{l}{\uppercase{Variables}} \\
\addlinespace[0.3em]
\quad \textit{n-agents} & 500$^*$ & 500$^*$ & 500$^*$ & 500$^*$\\
\quad \textit{pct-available-spots} & 22 & 22 & 22 & 22\\
\quad \textit{pct-whites} & Varies$^*$ & Varies$^*$ & Varies$^*$ & Varies$^*$\\
\quad $\textit{prob-unconditional-ia}_w$ & 0.10 & 0.10 & 0.10 & 0.10\\
\quad $\textit{prob-unconditional-sbe}_w$ & 0.11 & 0.11 & 0.11 & 0.11\\ 
\quad $\textit{prob-diffusion}_s$ & 1$^*$ & 1$^*$ & 1$^*$ & 1$^*$\\
\addlinespace[0.3em]
\multicolumn{5}{l}{\uppercase{Parameters}} \\
\addlinespace[0.3em]
\quad $\alpha_{\text{homophily}}$ & Eq.~\ref{eq:homophily}$^*$  & Eq.~\ref{eq:homophily}$^*$  &  Eq.~\ref{eq:homophily}$^*$  &  Eq.~\ref{eq:homophily}$^*$  \\
\quad $\alpha_{\text{n-ties}}$  & Eq.~\ref{eq:n-ties}$^*$  & Eq.~\ref{eq:n-ties}$^*$  & Eq.~\ref{eq:n-ties}$^*$   & Eq.~\ref{eq:n-ties}$^*$  \\
\quad $\alpha_{\text{acad-ineq}}$ & 1$^*$ & 1$^*$ & 1$^*$ & 1$^*$\\
\quad $\alpha_{\text{ia-ineq}}$ & 0.47 & 1$^*$ & Varies$^*$ & 0.47 \\
\quad $\alpha_{\text{sbe-ineq}}$ & 0.64 & 1$^*$ & Varies$^*$ & 0.64 \\
\quad $\alpha_{\text{favoritism}}$ & 0.9$^*$ & Varies$^*$ & 1$^*$ & Varies$^*$\\
\addlinespace[0.6em]
\hline
\end{tabular}
\end{table}

\FloatBarrier
\subsection{The emergence of Black-White advanced enrollment gaps across the share of White students in the school}

\begin{figure}[!t]
	\centering
	\includegraphics{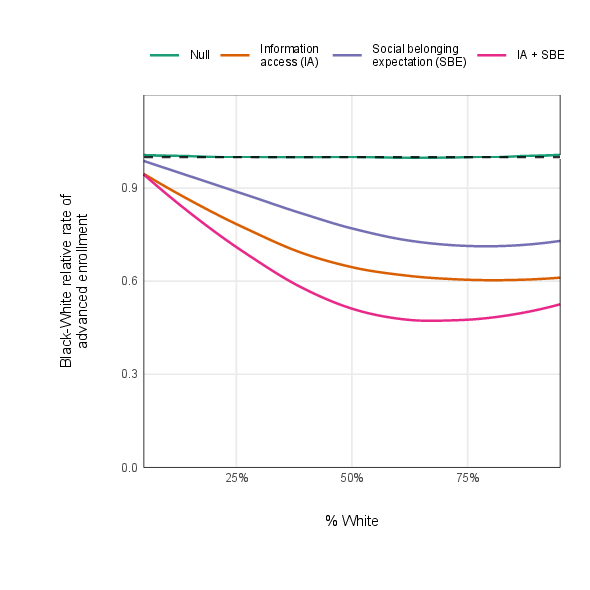}
	\caption{\textbf{The relationship between the Black-White relative rate of advanced enrollment and the share of White agents in school simulated by the agent-based model.}. Model variables and parameters initializing the model are defined by combination $C_1$ in Table~\ref{tab:initial-values}. Trending lines computed from 1,000 simulation runs for shares of White agents between 5\%-95\%. Results are presented for four different versions of the model: Null; IA; SBE; and IA + SBE, where the version name indicates the mechanisms shaping student placement. In the Null version, placement is only based on academic background.}
	\label{fig:fig-comp-pattern}
\end{figure}

Figure~\ref{fig:fig-comp-pattern} presents the relationship between the Black-White relative rate of advanced enrollment and the share of White agents in school simulated by the agent-based model.

Three points stand out. First, observe that the Null version of the model, on average, does not produce Black-White inequalities in advanced enrollment---$RR_{j})$ close to 1. This result is important to confirm that the model behaves as expected---i.e., when there are no Black-White inequalities in academic background, $\alpha_{\text{acad-ineq}} = 1$ and when selection is only based on academic requirements (Null version), no inequalities tend to emerge. It follows that any levels of Black-White inequalities which arise when other generative processes are introduced can be interpreted as inequalities between academically comparable Black and White students.

Second, the IA and SBE curves show that the dynamics of the IA and SBE mechanisms are similar, even though they are modelled in different ways.

Third, and most importantly, the IA, SBE and IA + SBE versions show that, on average, the inclusion of any of the mechanisms of interest (and of both) introduce a decreasing relationship between the Black-White relative rate of enrollment and the presence of White students in the high school. This means that the level of White advantages in advanced enrollment inequalities tends to increase as the presence of Whites in the school increases. Importantly, the complete model (IA + SBE) produces a relationship with the same central characteristics of the empirical pattern described in the literature (and illustrated in Figure~\ref{fig:fig-empirical-comp-pattern}). As in the empirical relationship, the simulated macro-pattern presents a decreasing and concave-up relationship: inequalities tend to increase as the presence of White students increases, but such increase only occurs up until White students represent about half of the student body, after this point, inequalities increase at much slower rate, and can even decrease.

These set of simulations, therefore, suggest that the presence of White students might shape the dynamics of competition for resources within schools and that such changing dynamics might constitute a plausible explanation for the well-known empirical pattern of interest. That said, given the many assumptions behind the model construction and the different moving-parts which might shape model outcomes, it is important to make sure emergent relationship is consistent with the theoretical foundations informing the model. To this end, therefore, we must understand exactly why simulations produce such macro-pattern. 

\FloatBarrier
\subsection{Understanding the emergent relationship}

To understand the dynamics of the model, note that agents' chances of advanced enrollment depend on two factors: their academic background ($\textit{prob-acad-qualification}_i$) and their eligibility for enrollment. Because of the assumption of no academic disparities between agents ($\alpha_{\text{acad-ineq}} = 1$), it follows that outcomes only depend on one's eligibility. Eligibility depends on meeting condition IA---i.e., having access to the information resources---and condition SBE---having optimistic social belonging expectations. If we understand social belonging expectations to also be a kind of \enquote{resource} shaping one's course-taking, then, we can say that one's eligibility depends on access to two kinds of resources: information and social expectations. The Black-White inequalities produced by the model, therefore, result from Black and White agents' unequal access to these two resources. 

Unequal access to these resources can arise in two different ways. First, it can arise exogenously to the model. Parameters can endow Black and White agents with initial (or structural) differences in access to the information and social expectations resources (parameters $\alpha_{\text{ia-ineq}}$ and $\alpha_{\text{sbe-ineq}}$, respectively). Second, it can arise endogenously to the model. Agents can have unequal opportunities for resource reception through social interactions (what is influenced by their network structure---parameters $\alpha_{\text{homophily}}$ and $\alpha_{\text{n-ties}}$---and by the extent to which same-race and different-race interactions differ---parameter $\alpha_{\text{favoritism}}$). Let us, now, carefully examine each of these sources of unequal access to the information and social expectations resources.

\FloatBarrier
\paragraph{Opportunities for resource reception through social interactions}

\begin{figure}[!t]
	\centering
	\includegraphics{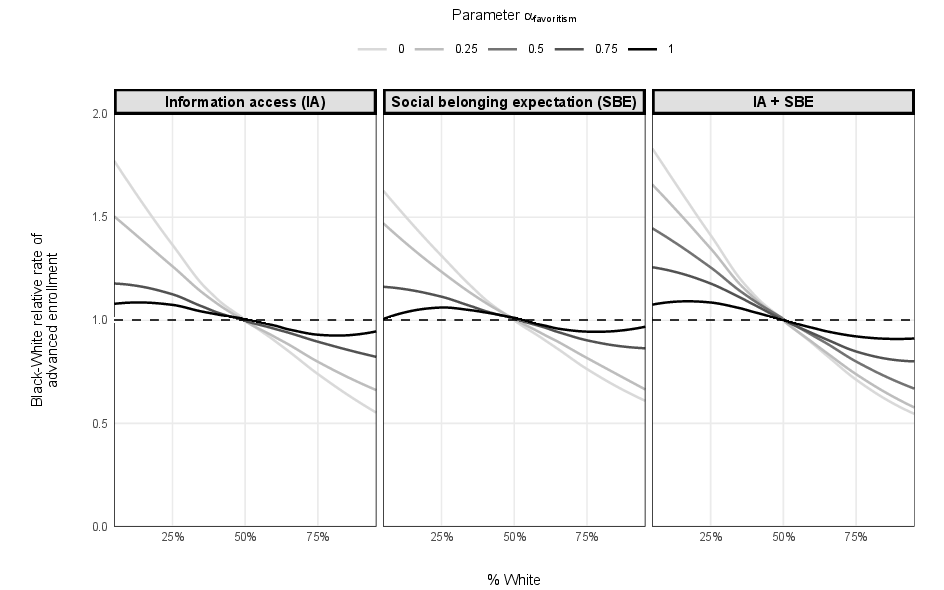}
	\caption{\textbf{The simulated relationship between the Black-White relative rate of advanced enrollment and the share of White agents in school for different values of parameter $\alpha_{\text{favoritism}}$.}. Model variables and parameters initializing the model are defined by combination $C_2$ in Table~\ref{tab:initial-values}. Trending lines computed from 100 simulation runs for shares of White agents between 5\%-95\%. Results presented for four different versions of the model: Null; IA; SBE; and IA + SBE, where the version name indicates the mechanisms shaping student placement. In the Null version, placement is only based on academic background.}
	\label{fig:fig-null-diffusion-gaps}
\end{figure}

To start unpacking the dynamics of the model, consider, first, the determinants of unequal opportunities for resource reception through social interactions. To do so, Figure~\ref{fig:fig-null-diffusion-gaps} plots simulation results for parameter/variable combination $C_3$ in Table~\ref{tab:initial-values}. In this combination, I assume away structural inequalities by setting $\alpha_{\text{ia-ineq}} = \alpha_{\text{sbe-ineq}} = \alpha_{\text{acad-ineq}} = 1$. Then, access to resources should only depend on the structure of agent's networks (parameters $\alpha_{\text{homophily}}$ and $\alpha_{\text{n-ties}}$) and on the extent to which same-race and different races interactions can differ (parameter $\alpha_{\text{favoritism}}$). To examine the role of these two features, combination $C_3$ allows $\alpha_{\text{favoritism}}$ to vary.

Note that when $\alpha_{\text{favoritism}} = 1$, there is no in-group favoritism in same-race interactions. Under this condition, therefore, the only factor shaping the dynamics of the model is the structure of agents' networks ($\alpha_{\text{homophily}}$ and $\alpha_{\text{n-ties}}$). Observe that the $\alpha_{\text{favoritism}} = 1$ curve produces some variations across the share of Whites in the school. Intuitively, this happens because the empirical calibration of networks imposes that number of network ties increases with relative group size (Fig~\ref{fig:fig-net}A), meaning that as the share of Whites in the school increases, the average number of ties for Whites increases while the average number of ties for Blacks decreases. Because agents can gain access to resources through social interactions with their ties, the more ties an agent has, the more opportunities for resource reception they have. Then, because parameter/variable combination $C_2$ imposes that opportunities for resource reception is the only factor shaping one's chances for advanced enrollment, it follows that chances for advanced enrollment should increase with relative size (what I call the \textit{relative size tendency}).\footnote{Note, for example, that, consistent with this \textit{relative size tendency}, no advanced enrollment inequalities emerge when both groups represent equal shares of the student body.}

\begin{figure}[!t]
	\centering
	\includegraphics{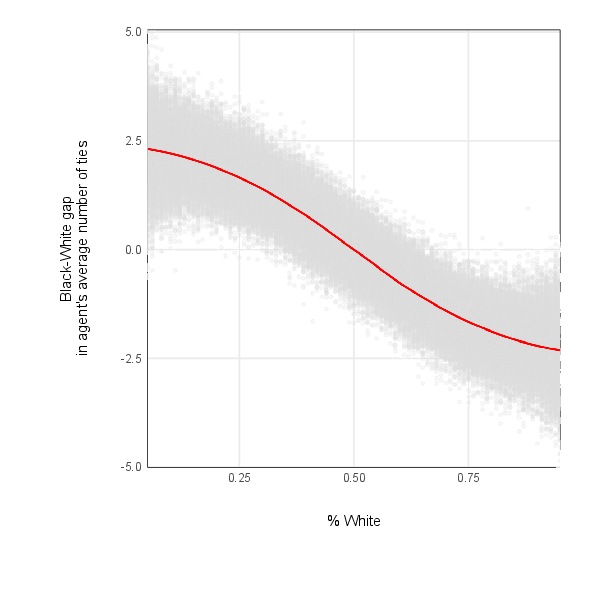}
	\caption{\textbf{The simulated relationship between the share of White agents in the model and gap in Black and White agents' average number of ties}. Trending line computed from 1,000 simulation runs for shares of White students between 1\%-99\%.}
	\label{fig:fig-n-ties-gap}
\end{figure}

To better visualize this \textit{relative size tendency}, Figure~\ref{fig:fig-n-ties-gap} plots the simulated Black-White gap in the average number of ties across the share of Whites in the school. Note that, in fact, the shape of this curve matches the shape of the $\alpha_{\text{favoritism}} = 1$ curve in Figure~\ref{fig:fig-null-diffusion-gaps}. 

Figure~\ref{fig:fig-null-diffusion-gaps} also illustrates the role of in-group favoritism on this \textit{relative size tendency}. When in-group favoritism increases (as $\alpha_{\text{favoritism}}$ gets closer to 0), the \textit{relative size tendency} is accentuated. To make sense of this result, note that when in-group favoritism plays a role in the model, agent's total number of ties is no longer the single factor shaping opportunities for resource reception. Because same-race ties are now more \enquote{valuable} than different-same ties, opportunities for resource reception must also depend on agent's number of same-ties. Then, because the structure of agents' networks imposes that the fraction of same-race ties increases with relative size (Fig~\ref{fig:fig-net}B), it follows that opportunities for resource reception---and, hence, also opportunities for advanced enrollment---will further increase with relative group, accentuating the \textit{relative size tendency} discussed above.

\FloatBarrier
\paragraph{Structural inequalities}

\begin{figure}[!t]
	\centering
	\includegraphics{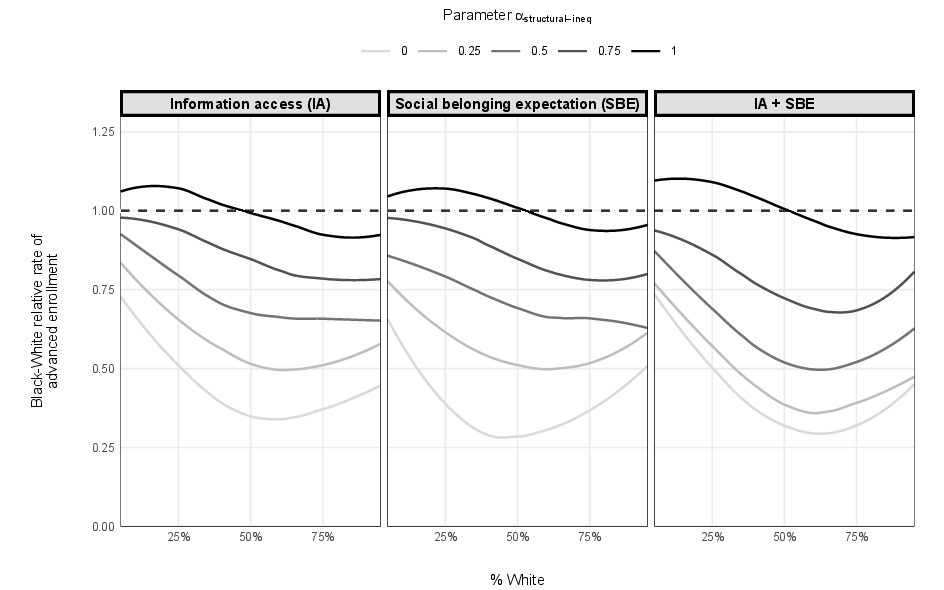}
	\caption{\textbf{The simulated relationship between the Black-White relative rate of advanced enrollment and the share of White agents in school for different values of parameter $\alpha_{\text{structural}}$, where $\alpha_{\text{structural}} = \alpha_{\text{ia-ineq}} = \alpha_{\text{sbe-ineq}}$.}. Model variables and parameters initialized from data as detailed in Table~\ref{tab:initial-values}, combination $C_3$. Trending lines computed from 100 simulation runs for shares of White agents between 5\%-95\%. Results presented for four different versions of the model: Null; IA; SBE; and IA + SBE, where the version name indicates the mechanisms shaping student placement. In the Null version, placement is only based on academic background.}
	\label{fig:fig-null-structural-gaps}
\end{figure}

Now, let us consider how the introduction of structural inequalities between Black and White agents might change these network-based tendencies. To do so, Figure~\ref{fig:fig-null-structural-gaps} plots simulation results for parameter/variable combination $C_3$ in Table~\ref{tab:initial-values}. In this combination, I allow structural inequalities between Black and White agents to vary. For simplicity, I set $\alpha_{\text{ia-ineq}} = \alpha_{\text{sbe-ineq}} = \alpha_{\text{structural-ineq}}$. Further, I assume way in-group favoritism ($\alpha_{\text{favoritism}} = 1$) and maintain the prior assumption of no differences in academic qualification ($\alpha_{\text{acad-ineq}} = 1$). 

The introduction of structural advantages for Whites leads to two noticeable patterns. First, it increases the White advantage for advanced enrollment across all shares of White in school---i.e., it shifts all curves downwards. Intuitively, this happens because inequalities at the start of the model lead to differences in how resourceful each race group's networks become---by setting structural advantages for Whites, for example, the model in Figure~\ref{fig:fig-null-structural-gaps} lead White networks to become more resourceful than Black's. When this happens, the rate at which Whites can gain access to resources becomes much faster than the rate at which Blacks can do so. Then, Whites can more quickly meet the conditions for advanced enrollment eligibility, being, generally, able to enroll in the advanced course before Blacks. When Blacks gain access to the resources necessary to meet eligibility conditions, most spots in the course are already full. 

However, a second pattern also emerges: the introduction of structural inequalities changes the shape of the resulting macro-level relationship, generating a tendency for curves to become concave up. Note, in fact, that the higher the structural inequalities, the more concave up the curve becomes. This happens because at extreme shares of Whites (either low are high), the rates of resource reception for Whites and Blacks tend to be more closely linked. Such closer link has to do with the fact that the fraction of different-race ties for each race group peaks when relative group size is low (see Figure~\ref{fig:fig-net}B). Then, at extreme shares of Whites, the group in the minority becomes closely integrated within the network of majority group. Therefore, even though the introduction of White structural advantages leads White networks to become more resourceful than Blacks', when the school is composed by extreme shares of Whites, Black-White differences in opportunities for resource reception cannot increase very much since one group's network is closely integrated within the other's.

\FloatBarrier
\paragraph{Making sense of the complete model}

Together, the combining roles of this network-based and structural-based tendencies help us make sense of the macro-level pattern observed in Figure~\ref{fig:fig-comp-pattern}. Given variable/parameter combination $C_1$ (Table~\ref{tab:initial-values}) defining the model in Figure~\ref{fig:fig-comp-pattern}, we can say that the structure of agent's networks together with some level of in-group favoritism, creates a \textit{relative size tendency}, that is, the tendency that opportunities for advanced enrollment should increase with relative size. Further, because combination $C_1$ introduces structural advantages for Whites over Blacks, it exacerbates the White advantages in opportunities for resource reception through social interactions, increasing the Black-White gap in advanced enrollment. That said, such structural advantages do not increase inequalities at the same rate when Whites represent extreme shares of the student body, which produces the observed decreasing and concave-up curves. 

Importantly, note that the IA + SBE curve tends to be more concave than the others. This happens because when both mechanisms play a role in the model, initial endowment with information resources is defined as a condition for initial endowment with social belonging expectations, which increases structural inequalities. The higher structural inequalities imposed in this model version, therefore, lead to a macro-pattern that is more concave-up, as suggested by the structural-based tendency illustrated in Figure~\ref{fig:fig-null-structural-gaps}.

\FloatBarrier
\section{Conclusions}
\label{sec:discussion}

This study set out to enhance our understanding of the mechanisms behind a well-known and important finding within the sociology of education literature, namely, that the presence of White peers correlates with an increase in the Black-White differential access to valuable school resources---defined as advanced coursework. Common explanations for this pattern face important limitations. Fist, scholars emphasize a cultural perspective which makes sense of this pattern through empirically questionable racial differences in peer culture \cite{fryer2010empirical}. Second, scholars propose a structural perspective which explains the macro-pattern through a third, confounding, factor: the historically-build underrepresentation of Black students in advanced courses in racially diverse and majority White schools \cite{tyson2011, francis2021separate}. Although this structural view provides a valuable contribution, it provides an explanation for the \textit{reproduction} (but not \textit{emergence}) of the macro-level pattern. Doing so, it does not consider the extent to which the presence of White peers might actually be a factor shaping Black-White within-school inequalities.

In this paper, through careful qualitatively informed and empirically based simulations, I show that we have reason to believe that the presence of White students in the school can actually influence the distribution of educational resources between Black and White students. In fact, I show that this influence can represent another possible explanation for the observed relationship between school composition and Black-White course-taking inequalities. It follows, therefore, that the common structural explanation for the macro-pattern of interest might be \textit{incomplete}, overlooking an important link between the presence of Whites and the emergence of school inequalities. This result provides useful insights on policy strategies to improve the experiences of Black students in racially diverse and majority-White schools, as discussed below.

First, this result illustrates that improving Black students' access to educational resources in majority-White and racially diverse schools might be more complex than what the common structural view suggests. If interpreted as a complete explanation for the empirical pattern of interest, this structural view suggests that closing historically built gaps in advanced enrollment should interrupt a cycle of social reproduction, and, ultimately, could prevent higher inequalities to arise in the presence of more White students. In contrast, because the simulation-based investigations presented here suggest that the presence of White students can actually lead to the emergence (and not simply reproduction) of higher resource-access inequalities, it follows that closing the historical underrepresentation of Black students in advanced courses would not, by itself, eliminate the inequality-producing tendencies brought about by the presence of White students. Therefore, efforts to improve Black students' access to advanced courses in majority-White/racially diverse schools should also consider the characteristics of the student selection process which allow the presence of Whites to bring out course-taking inequalities.

Second, by highlighting some of the mechanisms through which such inequalities emerge, this study suggests possible avenues for intervention. From the model here, we learn that higher levels of inequality in arise in the presence of White students because of the role informal resources (such as information resources and social belonging expectations) play in the advanced enrollment process. It follows, therefore, that interventions which attempt to equalize the advanced enrollment opportunities of similarly qualified Black and White students can be promising if they reduce the value these informal resources have to students' advanced placement chances. For example, to reduce the salience of information resources to one's enrollment chances, it could be useful to make the course-taking process easier to navigate, establishing a unified, transparent, enrollment criteria that is consistent across public schools, which contrasts, for instance, the strong school-level variation which defines current practices \cite{kelly2007contours}. Similarly, the informed practices of students (and of their families) could be less of an advantage if placement decisions were less dependent on the recommendation of teachers, recommendations which, as discussed, are often shaped by the pressure of highly involved parents and students \cite{lewis2015}. Further, by recognizing that students' motivation for advanced courses is shaped by their networks and not their cultural dispositions, it makes sense to counter common practices which emphasize students' motivation when making placement decisions---e.g., the current notion that advanced placement (AP) courses are reserved to highly academically motivated students \cite{CollegeBoard-AP-courses}. Rather, it is useful to attribute to schools the role of influencing the motivation of all academically qualified students. 

Beyond its contribution to our understanding of Black-White within-school inequalities, this study, more generally, brings valuable methodological insights to research on the organization of schools. Because the internal dynamics of schools depend on the decisions of and social interactions between multiple actors---such as students, families, administrators, and school staff--—understanding the organizational behavior of schools can be a challenge \cite{lewis2014inequality, mcfarland2014network, frank2018social, small2009unanticipated}. Given this complexity, an understanding of the mechanisms shaping the educational experiences of students can benefit from agent-based modeling approaches, which allow us to more systematically analyze the social-dependency of human behavior. This approach is particularly promising in education research given the vast availability of qualitative foundations concerning the behaviors of teachers, parents and students. Such rich literature, therefore, can provide an empirical foundation to address a central step in agent-based modeling explorations: the design of reasonable computational rules governing agent behavior, allowing models to better contribute to our understanding of the internal dynamics of schools.

\FloatBarrier

\section*{Statements and Declarations}

\subsection*{Acknowledgements}
I thank R. L’Heureux Lewis-McCoy, Samuel Lucas, Erez Hatna, Ravi Shroff, Gianluca Manzo, John Skvoretz, Lisa Stulberg and Yasmiyn Irizarry for helpful comments and feedback. Also, I note that an earlier version of this paper was presented at the 2022 American Sociological Association Annual Meeting, Session on Mathematical and Computational Approaches to Studying Inequality (August 2022) and at the The Seventh Joint US-Japan Conference on Mathematical Sociology and Rational Choice (August 2022).

\subsection*{Conflict of interest:} The authors declare that they have no competing interests.

\subsection*{Funding:} The authors acknowledge that they received no funding in support for this research.

\subsection*{Data availability:} The data used in this paper results from a combination of publicly available datasets: the 2017-18 Common Core of Data, from the National Center for Education Statistics; the 2017-18 Civil Rights Data Collection, also from the National Center for Education Statistics; and the the Stanford Education Data Archive version 4.0---https://doi.org/10.25740/db586ns4974---\cite{seda}. 

\subsection*{Materials availability:} The code to fully reproduce all results in this paper is available from the corresponding author on reasonable request and will be made publicly available upon publication of this and subsequent related projects.

\pagebreak
\begin{singlespacing}
\bibliography{references-school-effects.bib,references-course-taking.bib,references-peer-effects.bib,references-discrimination.bib, references-ABM.bib, references-opportunity-hoarding.bib}
\bibliographystyle{asr}
\end{singlespacing}
\renewcommand{\bibname}{References}

\pagebreak
\appendix
\renewcommand{\thesection}{Appendices}

\renewcommand\thefigure{\thesection.\arabic{figure}} 
\renewcommand\thetable{\thesection.\arabic{table}} 
\setcounter{figure}{0}  
\setcounter{table}{0}  

\FloatBarrier
\renewcommand{\thesection}{Appendix A}
\section{Sample and variable selection for model initialization} 
\label{A:sample}
\renewcommand{\thesection}{A}

I rely on data from the 2017-2018 Common Core of Data (CCD), the 2017-2018 Civil Rights Data Collection (CRDC) and the Stanford Education Data Archive version 4.0 \cite{seda}, to examine Black and White students' enrollment in advanced placement (AP) mathematics courses. I focus specifically on high schools that (1) offer grades 9-12; (2) are defined by the CCD to be \enquote{regular schools} (instead of special education, vocational, alternative or reportable program schools); (3) were operational in 2017 and 2018; and (4) offer AP courses but not International Baccalaureate (IB) courses---since IB availability might change the dynamics of AP enrollment.

Because students tend to take AP mathematics primarily in grade 12 \cite{judson2015growth}, I narrow my attention to the cohort of students who were in \nth{12} grade during the 2017-18 academic year.\footnote{Although CRDC's measure of AP course-taking compute the total number of high schoolers who enrolled in at least one AP math course in the given academic year, dividing such total number by all students in the high school will likely underestimate the rate at which students take AP math courses in their high school careers as students in grades 9 or 10 rarely take AP math courses \cite{judson2015growth}. Further, even if a student did not take any AP math courses in 11 grade, they could still take courses in \nth{12} grade before they graduate. It follows, therefore, that to capture the extent to which students take at least some AP math during their high school experiences, it is useful to use the number of \nth{12} grade students as the denominator---a strategy further supported given that those who do take some AP math in \nth{11} are likely to take some AP math in \nth{12} grade as well.} Further, given the focus on the dynamics between Black and White students, I narrow the sample to high schools where Black and White students represent at least 75\% of the \nth{12} grade cohort. Finally, to focus on schools which offer AP mathematics courses, I narrow the sample to those in AP mathematics enrollment is higher than 5\%. Together, these restrictions lead to a sample of 3,482 high schools. This is the sample which I use for the analysis in Figure~\ref{fig:fig-empirical-comp-pattern}.

Now, I discuss the mapping of empirical variables into model variables. Table~\ref{tab:data-sources} details the construction of all empirical variables of interest.

\begin{itemize}
    \item Variables \textbf{\textit{number-agents}, \textit{pct-whites} and \textit{pct-blacks}} are initialized, respectively, with empirical variables \textit{SCH: AP math enrollment rate (Blacks + White)}, \textit{Pct. White} and \textit{Pct. Black}. Note that the model only allows agents to be either Black or White. Then, I calculate empirical variables assuming that schools are composed by Black and White students only. Therefore, \textit{N. 12 grade (Black + White)} is the total of number of students in the \nth{12} grade cohort which are Black or White. Further, \textit{Pct. White} and \textit{Pct. Black} are calculated with \textit{N. 12 grade (Black + White)} in the denominator so that $\textit{Pct. White} + \textit{Pct. Black} = 100\%$.
    \item Variable \textbf{\textit{pct-available-spots}} is initialized with empirical variable \textit{AP math enrollment rate (Blacks + White)}.
    \item Parameter \textbf{$\alpha_{\text{acad-ineq}}$} is initialized with the fraction of \textit{Pct. algebra 1, MS (Blacks)} over \textit{Pct. algebra 1, MS (Whites)}. These empirical variables represent Algebra 1 course-taking patterns in the district's middle schools. Note that the completion of Algebra 1 before high school is a well-known predictor of an advanced mathematics course-taking trajectory \cite{francis2021separate} and, thus, an useful indication of students' academic background. Note, also, that to provide an indication of the academic background of the cohort of students, I construct middle school variables using 2013-14 data. Assuming ideal grade promotion and no transfers across districts, this is the year in which students in the cohort of interest were in \nth{8} grade. 
    \item Variables/parameter \textbf{$\textit{prob-unconditional-ia}_W$, $\textit{prob-unconditional-ia}_B$, $\alpha_{\text{ia-ineq}}$} are initialized, respectively, with $(\textit{Pct. parents with BA+ (Whites)} / 100)^2$, \\ $(\textit{Pct. parents with BA+ (Whites)} / 100)^2$ and $\frac{(\textit{Pct. parents with BA+ (Blacks)}/ 100)^2}{(\textit{Pct. parents with BA+ (Whites)}/100)^2}$. These model variables capture unconditional endowment with information resources. From qualitative research, unconditional access to these resources often results from parents' ability to successfully assist and coordinate their children's education---an ability correlated with parents' social class and educational backgrounds \cite{bodovski2008concerted, lareau2011}. Thus, I initializ these variables based on indicators of parental educational background. Importantly, such kind of parental involvement is rare, and seems to be particular to those in the very high-end of the social class distribution \cite{calarco2018, lareau2011, lewis2014inequality}. Then, to capture how selective such parental practices are, I elevate the indicators of educational background to the power of 2.
    \item Variables/parameter \textbf{$\textit{prob-unconditional-sbe}_W$, $\textit{prob-unconditional-sbe}_B$, $\alpha_{\text{sbe-ineq}}$} are initialized, respectively, with $0.5 \cdot \textit{Pct. algebra 1, MS (Whites)}$, $0.5 \cdot$ \textit{Pct. algebra 1, MS (Blacks)} and $\frac{\textit{Pct. algebra 1, MS (Blacks)}}{\textit{Pct. algebra 1, MS (Whites)}}$. These model variables capture students whose social belonging expectations are automatic---i.e., unconditional to the messages they receive in school. From qualitative research, these students are the ones with very high academic backgrounds. Then, I also initialize these variables based on empirical indicators of academic background. That said, to emphasize that such unconditional feelings of belonging are rare, and that even students who come from strong academic backgrounds might depend on the messages they receive from their peers \cite{oconnor2011being}, I multiply these variables by 0.5.
\end{itemize}

Given this empirical initialization of model variables, further restrictions to the sample of interest are useful. I emphasize that by introducing these further restrictions, I am not concerned with achieving a nationally representative sample. Rather, the goal here is to select schools which can ground model variables on some empirical reality, avoiding the use of purely hypothetical distributions.  First, I restrict the sample to schools with at least 10 Black and 10 White students in the \nth{12} grade cohort. Second, because the model only contains Black and White agents, I concentrate on schools where the AP mathematics enrollment rate for Black and White students (combined) is higher than 5\%. Third, because some of my measures used for model initialization are calculated at the district-level, I narrow my analysis to school districts with only one high school institution---which, presumably, can increase the reliability of these measures at the high-school level. Finally, Algebra 1 enrollment in grades 7 and 8 should be non-zero for Blacks and Whites (combined) and the share of parents with at least a BA degree should be non-zero for Blacks and Whites (combined).\footnote{Variables must be non-missing, which removes 2 high schools from the sample.} Together, these filters total to 530 high schools which are used for model initialization.

\begin{sidewaystable}[!t]
	\footnotesize
	\centering
	\singlespacing
	\caption{\textbf{Empirical variables for model initialization.}}
	\label{tab:data-sources}
 \begin{threeparttable}
\begin{tabular}{p{7cm}p{10cm}p{1cm}wc{1cm}wc{1cm}}
\toprule
\addlinespace[0.3em]
\text{Empirical variable} & Details & Source & Mean & S.d.\\
\addlinespace[0.3em]
\midrule
\addlinespace[0.6em]
\multicolumn{3}{l}{\uppercase{School characteristics}}\\
\addlinespace[0.3em]
\hspace{1em}N. 12 grade (Black + White) & Number of Black plus White students in grade 12. & CCD & 251.8 & 136.0\\
\hspace{1em}Pct. White & Number of Whites in grade 12 divided by the number of Blacks plus Whites in grade 12. & CCD & 19.2 & 18.2	\\
\hspace{1em}Pct. Black & Number of Blacks in grade 12 divided by the number of Blacks plus Whites in grade 12. & CCD & 80.8 & 18.2\\
\addlinespace[0.3em]
\multicolumn{3}{l}{\uppercase{Academic preparation}}\\
\addlinespace[0.3em]
\hspace{1em}Pct. algebra 1, MS (Whites) & Percentage of \nth{7} and \nth{8} grade White students in the district who enrolled in Algebra 1 during the 2013-14 academic year. & CRDC & 21.2 & 13.2\\
\hspace{1em}Pct. algebra 1, MS (Blacks) & Percentage of \nth{7} and \nth{8} grade Black students in the district who enrolled in Algebra 1 during the 2013-14 academic year. & CRDC & 13.7 & 14.5	\\
\addlinespace[0.3em]
\multicolumn{3}{l}{\uppercase{Parental background}}\\
\addlinespace[0.3em]
\hspace{1em}Pct. parents with BA+ (Whites) & District-level share of White families with and educational background that is equal to or higher than a bachelor's degree. & SEDA & 32.4 & 15.7\\
\hspace{1em}Pct. parents with BA+ (Blacks) & District-level share of Black families with and educational background that is equal to or higher than a bachelor's degree. & SEDA & 22.2 & 12.7\\
\addlinespace[0.3em]
\multicolumn{3}{l}{\uppercase{AP mathematics course-taking}}\\
\addlinespace[0.3em]
\hspace{1em}AP math enrollment rate (Whites) & Number of White students enrolled in at least one AP mathematics course over all White students in grade 12. & CRDC & 24.8 & 15.7\\
\hspace{1em}AP math enrollment rate (Blacks) & Number of Black students enrolled in at least one AP mathematics course over all Black students in grade 12& CRDC & 8.3 & 9.4\\
\hspace{1em}AP math enrollment rate (Blacks + Whites) & Number of White plus Black students enrolled in at least one AP mathematics course over all White and Black students in grade 12. & CRDC & 21.6 & 13.8	\\
\addlinespace[0.6em]
\bottomrule
\end{tabular}
\begin{tablenotes}
     \tiny 
     \item[1.] CCD values reported for the 2017-18 academic year.
     \item[2.] CRDC values reported for the 2017-18 academic year.
     \item[3.] SEDA values reported as an estimated average for academic years between 2009-18.
\end{tablenotes}
\end{threeparttable}
\end{sidewaystable}

\pagebreak
\FloatBarrier
\setcounter{figure}{0}  
\setcounter{table}{0}  
\renewcommand{\thesection}{Appendix B}
\section{Notes on the network calibration process}
\label{A:net}
\renewcommand{\thesection}{B}
\FloatBarrier
The goal of the network formation model is to endow agents with friendship ties which follow findings from the social network literature, namely, that the tendency of forming same-race ties varies with the racial composition of schools, peaking in racially heterogeneous settings \cite{mouw2006residential, moody2001race, currarini2010identifying}. The choice of \citet{currarini2010identifying} over other studies for model calibration is due to the amount of technical details provided by the authors, providing multiple tools for calibrating and evaluating the results of the ABM here.

\begin{figure}[h]
	\centering
	\singlespacing
	\includegraphics{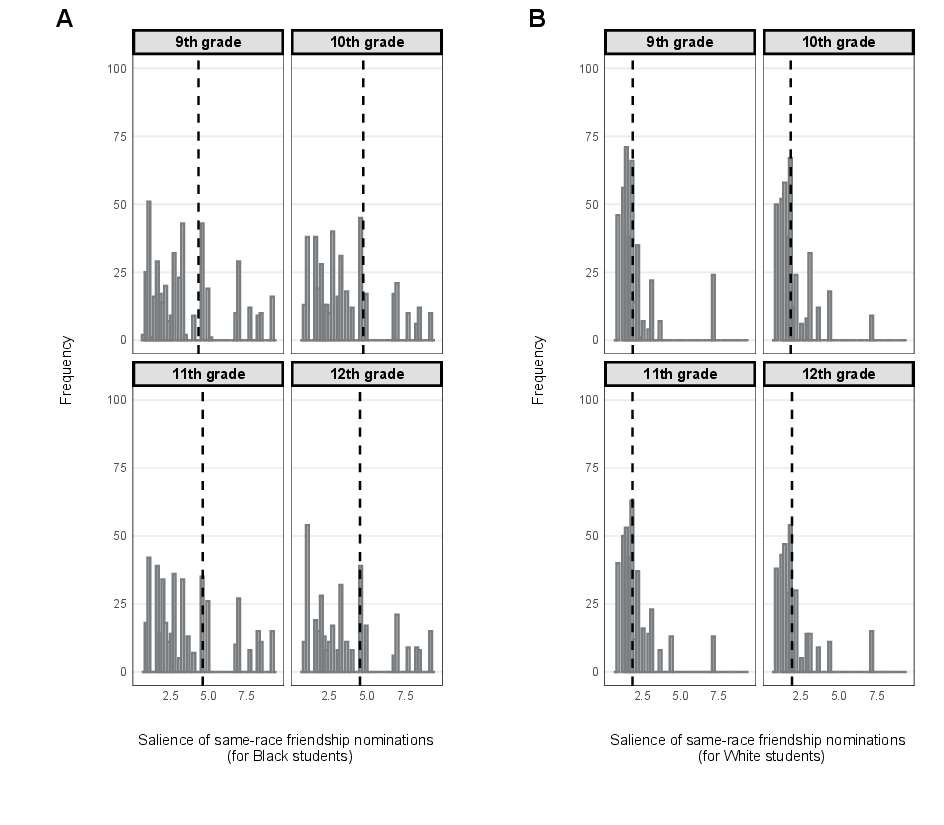}
	\caption{\textbf{Histograms of the salience of same-race ties in students' friendship networks students across high school grades (9-12), computed from the National Longitudinal Study of Adolescent to Adult Health, Add Health, \cite{addhealth}.} \textbf{Plot A} provides results for Black students and \textbf{Plot B} provides results for White students. The salience index is a measure of the extent to which students tend to nominate same-race students as friends. It is calculated by the Add Health data according to \cite{rytina1982arithmetic}.}
	\label{fig:Add-Health}
\end{figure}

Note that because the model is initialized to represent the context of AP mathematics enrollment, students' networks should, ideally, reflect friendship patterns before AP mathematics placement decisions take place, which is often around the end of students' \nth{10} and/or \nth{11} grades \cite{judson2015growth}. Although \citet{currarini2010identifying} rely on data from students in all high school grades---as provided by the National Longitudinal Study of Adolescent Health (Add Health)---their results can be seen as a reasonable approximation of students' friendship patterns before AP mathematics decisions take place since a descriptive investigation of the Add Health data shows that the salience of same-race friendship ties for Black and White students is virtually unchanged across high school grades (see Figure~\ref{fig:Add-Health}).

\pagebreak
\FloatBarrier
\setcounter{figure}{0}  
\setcounter{table}{0}  
\renewcommand{\thesection}{Appendix C}
\section{Sensitivity analysis for in-group favoritism parameter}
\label{A:sens}
\renewcommand{\thesection}{C}

\begin{figure}[t]
	\centering
	\includegraphics{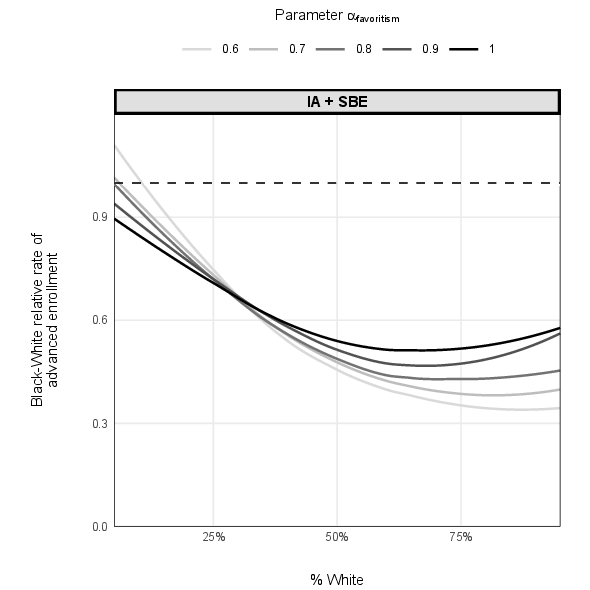}
	\caption{\textbf{The simulated relationship between the Black-White relative rate of advanced enrollment and the share of White agents in school for different values of parameter $\alpha_{\text{favoritism}}$.}. Model variables and parameters initializing the model are defined by combination $C_4$ in Table~\ref{tab:initial-values}. Trending lines computed from 100 simulation runs for shares of White agents between 5\%-95\%. Results presented for four different versions of the model: Null; IA; SBE; and IA + SBE, where the version name indicates the mechanisms shaping student placement. In the Null version, placement is only based on academic background}
	\label{fig:fig-sens-diffusion-gaps}
\end{figure}

Parameter $\alpha_{\text{favoritism}}$ is not initialized from data. I choose its value based on three principles. First, consistent with the empirical notion that agents often favor interactions with same-race peers,  $\alpha_{\text{favoritism}}$ must be lower than 1. Second, as discussed in the model validation section, when the model is applied to the empirical context of interest (\ref{A:sample}), outcomes should closely reproduce, but not overestimate, the levels of Black-White inequalities observed in the data. Table~\ref{tab:sens-diffusion} shows the average model error, $ME$, across the 530 high schools of interest for different values of the $\alpha_{\text{favoritism}}$ parameter. Results show that values between 0.7 and 1 can satisfy this principle. Finally, because part of the goal of the analysis is to reproduce the macro-level pattern described in Figure~\ref{fig:fig-empirical-comp-pattern}, I select the parameter value which allows the model to best mimic this relationship. Figure~\ref{fig:fig-sens-diffusion-gaps} presents the simulated macro-pattern which emerges from different values of $\alpha_{\text{favoritism}}$. Plots show that higher levels of $\alpha_{\text{favoritism}}$ allow the model to be more concave-up. Then, balancing these three principles of interest, I choose $\alpha_{\text{favoritism}} = 0.9$.

\begin{table}[h]
        \footnotesize
	\centering
	\caption{\textbf{Variations in model performance across values of parameter $\alpha_{\text{favoritism}}$.} Model performance computed based on model's error rate \textit{ME}---the average difference between the true and simulated values across all high schools in the empirical context of interest (\ref{A:sample}). In the model version considered, student placement is shaped by both the IA and SBE mechanisms.}
	\label{tab:sens-diffusion}
	\begin{tabular}{l | c }
\hline
\addlinespace[0.3em]
 $\alpha_{\text{favoritism}}$ & $ME$ \\
 \addlinespace[0.3em]
 \hline
\addlinespace[0.3em]
0.7 &  -0.0224\\
0.8 &  -0.668\\
0.9 &  -1.31\\
1  &  -1.92\\
\addlinespace[0.6em]
\hline
\end{tabular}
\end{table}
\FloatBarrier

\end{document}